\documentclass[3p]{elsarticle}
\usepackage{aasmacros}
\usepackage{amsmath, amssymb}
\usepackage{multicol}
\usepackage{xcolor}
\usepackage{xspace}
\usepackage{enumerate}
\usepackage{tablefootnote}
\usepackage{threeparttable}
\usepackage{booktabs}
\usepackage[english]{babel}
\usepackage[section]{placeins}
\usepackage[pdfpagelabels, plainpages=false, linktocpage=true]{hyperref}

\hypersetup{
   colorlinks   = true, 
   urlcolor     = blue, 
   linkcolor    = MidnightBlue, 
   citecolor    = red 
}

\journal{PoS: 38th ICRC 2023}

\begin{document}

\title{CRPropa 3.2: a public framework for high-energy astroparticle simulations}

\author[1,2]{S.~Aerdker}
\author[3]{R.~Alves Batista}
\author[1,2,4]{J.~Becker Tjus}
\author[1,2]{J.~Dörner}
\author[5,6]{A.~Dundovic}
\author[1,2]{B.~Eichmann}
\author[1,2]{A.~Frie}
\author[7,8]{C.~Heiter}
\author[1,2,9]{M.~Hoerbe}
\author[10,2]{K.-H.~Kampert}
\author[1,2]{L.~Merten\corref{cor1}}
\ead{crpropa@desy.de}
\author[7]{G.~Müller}
\author[9]{P.~Reichherzer}
\author[11]{S.~Rossoni}
\author[12,13]{A.~Saveliev}
\author[1,2]{L.~Schlegel}
\author[11]{G.~Sigl}
\author[14]{A.~van Vliet} 
\author[8]{T.~Winchen}

\affiliation[1]{Theoretical Physics IV, Faculty for Physics \& Astronomy, Ruhr University Bochum, Germany}
\affiliation[2]{Ruhr Astroparticle and Plasma Physics Center (RAPP Center), Germany}
\affiliation[3]{Instituto de Física Teórica UAM-CSIC, Madrid, Spain}
\affiliation[4]{Department of Space, Earth and Environment, Chalmers University of Technology, Gothenburg, Sweden}
\affiliation[5]{Gran Sasso Science Institute (GSSI), L’Aquila, Italy}
\affiliation[6]{Institute for Cosmology and Philosophy of Nature (ICPN), Kri\v{z}evci, Croatia}
\affiliation[7]{RWTH Aachen University, III. Physikalisches Institut A, Aachen, Germany}
\affiliation[8]{Max Planck Institute for Radio Astronomy, Bonn, Germany}
\affiliation[9]{Department of Physics, University of Oxford, Oxford, United Kingdom}
\affiliation[10]{Bergische Universität Wuppertal, Department of Physics, Wuppertal, Germany}
\affiliation[11]{II. Institute for Theoretical Physics, Universit\"at Hamburg, Hamburg, Germany}
\affiliation[12]{Immanuel Kant Baltic Federal University, Institute of Phys., Math.\ and Inf.\ Techn., Kaliningrad, Russia}
\affiliation[13]{Lomonosov Moscow State University, Faculty of Comp.\ Mathematics and Cybernetics, Moscow, Russia}
\affiliation[14]{Department of Physics, Khalifa University, Abu Dhabi, United Arab Emirates}

\cortext[cor1]{Presenter}

\begin{abstract}
CRPropa is a Monte Carlo framework for simulating the propagation of (ultra-) high-energy particles in the Universe, including cosmic rays, gamma rays, electrons, and neutrinos. It covers energies from ZeV down to GeV for gamma rays and electrons, and TeV for cosmic rays and neutrinos, supporting various astrophysical environments such as the surroundings of astrophysical sources, galactic, and extragalactic environments. The newest version, CRPropa 3.2, represents a significant leap forward towards a universal multi-messenger framework, opening up the possibility for many more astrophysical applications. This includes extensions to simulate cosmic-ray acceleration and particle interactions within astrophysical source environments, a full Monte Carlo treatment of electromagnetic cascades, improved ensemble-averaged Galactic propagation, significant performance improvements for cosmic-ray tracking through magnetic fields, and a user-friendly implementation of custom photon fields, among many more enhancements. This contribution will give an overview of the new features and present several applications to cosmic-ray and gamma-ray propagation.
\end{abstract}


\maketitle

\section{Introduction}
In the light of recent multi-wavelength and sometimes even multi-messenger observation of astrophysical objects a detailed modelling of transport and interaction of cosmic rays and their secondary particles is vital. Since its last major release \citep{CRPropa3} the simulation framework CRPropa has made progress to become an all particle (almost) all energies modelling tool. Since version 3.1 \citep{CRPropa31} CRPropa can not only propagate cosmic-ray \texttt{Candidates} by solving their equation of motion but also solve an ensemble-averaged Fokker Planck transport equation based on a stochastic differential equation (SDE) solver.

Ever since, CRPropa's modular structure makes improving and extending the existing software easy. In these proceedings we report about the latest developments of the code that have been implemented since the last major release version 3.2 and the publication of the corresponding paper \cite{CRPropa32}. This work is structured as follows: In section \ref{sec:tec} technical improvements, such as \texttt{tagOrigin} and \emph{plugins}, are introduced. New physics use cases and example simulations are discussed in section \ref{sec:phys}. And outlook into the future of CRPropa is given in section \ref{sec:outlook}.

\section{Technical Improvements}
\label{sec:tec}

\subsection{Improved Propagation Algorithms}
\label{ssec:propalgos}
CRPropa provides three different propagation algorithms to solve the equation of motion \texttt{SimplePropagation}, \texttt{PropagationBP}, and \texttt{PropagationCK}. For comparisons of the latter two, the reader is referred to \cite{Reichherzer20, CRPropa32} and the documentation\footnote{\url{https://crpropa.github.io/CRPropa3/pages/example\_notebooks/propagation\_comparison/Propagation\_Comparison\_CK\_BP.html}}. Both solvers are implemented adaptively by default. This means the module optimises the propagation step $\Delta s$ within a user-defined range $\Delta s \in [s_\mathrm{min}, s_\mathrm{max}]$ to match a given precision. Usually this gives an overall performance improvement compared to non-adaptive algorithms because the integration step will be increased, e.g., in regions of low magnetic field strength. When transport properties such as the diffusion coefficients should be studied it is, however, often advantageous to use a fixed propagation step in the simulation, making the usual error estimation and step optimisation obsolete. Now the corresponding computations are skipped when \texttt{PropagationBP} and \texttt{PropagationCK} are initialised with the values for the minimal and maximal propagation step $s_\mathrm{min}=s_\mathrm{max}$, leading to performance improvements of approximately a factor 2-3.\footnote{See (\url{https://github.com/CRPropa/CRPropa3/pull/424}) for details on the performed tests.} The computation time of the adaptive versions remains the same, as well as all simulation results. 

\subsection{Tracking of Secondary Production}
\label{ssec:interactiontag}
Although CRPropa tracks individual particles it was up to now not possible to differentiate particles due to their production mechanism. With the introduction of a new \texttt{Candidate} property called \texttt{tagOrigin}
the producing interaction is now recorded for all particles. This allows to separately analyse observables, such as secondary spectra, by their origin. This will help to better understand the dominating production mechanism. Furthermore, this feature can be used to identify potentially irrelevant interactions. When these are removed from the model, significant simulation speed-ups are expected. 

By default \texttt{tagOrigin}s are based on the kind of interaction, e.g. photo-pion production and inverse Compton scattering, but do not differentiate between different photon target fields. This behaviour can be changed by providing individual tags for each module instance before adding it to the \texttt{ModuleList}. Furthermore, the new feature can be used to differentiate between different source classes in post-processing as it can be modified by \texttt{Sources}, too. 

Recording the \texttt{tagOrigin} leads to a small performance overhead of approximately three percent for simulations that are dominated by interactions.\footnote{See (\url{https://github.com/CRPropa/CRPropa3/pull/407}) for details.} Note, that old post-processing scripts might break since the \texttt{tagOrigin} is now included in some of the predefined output formats.

\subsection{Extended Density Grids}
\label{ssec:grids}
In version 3.2 the new property of \texttt{Density} fields has been introduced. It was implemented as a first step towards nuclei-nuclei interactions, where they will serve as the description of the target fields. Here, we discuss two upgrades to this feature. Firstly, density grids are now restricted to their original volume defined by the origin and the number of and distance between grid points. Outside of this volume the fields are vanishing. The behaviour can be changed by the user to use reflective or periodic continuation as often done for turbulent magnetic fields. Note, that in general these new features allows to restrict other fields to their original volume, too.

Secondly, a new source class \texttt{SourceMassDistribution} was implemented. It allows to sample source positions from any given \texttt{Density} class, e.g., sampling cosmic-ray sources that follow one of the predefined gas distributions of the Galaxy.

\subsection{Customisation with Plugins}
\label{ssec:plugins}
CRPropa plugins are a way to provide extensions to CRPropa that are not planned to become part of the core CRPropa software package. Similar to the main CRPropa software plugins are written in C++ and wrapped for python access with SWIG. After compilation they can be imported as separate python modules and be freely combined with the CRPropa software. The documentation includes now an updated example, which can be used as a starting point for user developments. Although plug-ins are not maintained by the CRPropa developers we encourage users to share their developments with the community.

Plugins that have been shared are linked on the CRPropa. Some of them are explained exemplary in the following:

\paragraph{Field Line Integration}

The \texttt{FieldLineIntegrator} module can be used to analyse magnetic field models that are implemented in CRPropa. The module should be used instead of a conventional propagation module, e.g., \texttt{PropagationBP}, and will integrate the magnetic field line passing through the source position of the candidate. A simulation set-up to integrate magnetic field lines must not include interaction modules, but can include boundary conditions as the maximum trajectory length. In post-processing candidate positions can be used to, e.g., visualise a magnetic field model, calculate diffusion coefficients in non-homogeneous background fields, or field line random walk coefficients.

\paragraph{Plasma Instabilities}

Electron-positron pairs propagating in the intergalactic medium produce plasma instabilities that can affect their propagation. The exact role played by this effect is not known, but it can be sizeable and influence studies that depend on accurate modelling of electromagnetic cascades, such as constraints on intergalactic magnetic fields~\cite{alvesbatista2021a}.
The \texttt{GRPlInst} plugin\footnote{\url{https://github.com/rafaelab/grplinst}} was developed to take this into account. It interfaces with CRPropa and effectively adds a continuous energy-loss term to electrons and positrons, depending on the type of plasma instability and properties of the intergalactic medium~\cite{alvesbatista2019g}.

\paragraph{Lorentz Invariance Violation}

The recent interest in probing quantum gravity (QG) via multi-messenger astronomy \cite{addazi2022} was the main motivation behind the creation of the \texttt{LIVPropa} plugin \cite{ICRCLIV2023}. It incorporates the main phenomenological consequence of QG-based Standard Model extensions, namely the modification of the dispersion relations of different particles. Based on that, the propagation parameters are recalculated and can then be used in CRPRopa to evaluate the impact of these modifications on the observations. So far only pair production for photons with LIV is available, but we are planning to extend it to other reactions, such as inverse Compton scattering, soon.

\subsection{Customisation with Photon Fields}
\label{ssec:customphotonfields}
Although custom photon fields are technically available since version 3.2 a consistent implementation of a user-defined target photon fields was quite complicated. To make the procedure of adding a custom photon field clear, we first elaborate how interactions with photon fields are technically implemented in CRPropa. 

Since an on-the-fly derivation of all relevant interaction kinematics would be too time consuming, CRPropa uses pre-calculated tables for parameters, such as the interaction rate, which are already integrated over the target photon fields. To generate these tables a collection of python modules has been developed that can be found in a separate git repository\footnote{\url{https://github.com/CRPropa/CRPropa3-data}}. In addition, some CRPropa modules require direct access to photon field properties such as the number density. This means any new photon field has to be implemented twice: On the one hand, as a class of the CRPropa3-Data repository to allow for the generation of the tabulated interaction rates and on the other hand as CRPropa class to allow for direct access. 

\begin{figure}[htbp]
    \centering
    \includegraphics[width=.65\textwidth]{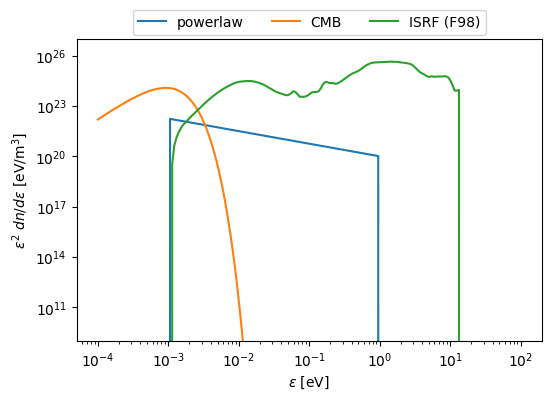}
    \caption{Examples of two custom photon fields in comparison with the cosmic microwave background.}\label{fig:customphotonfields} 
\end{figure}

However, the actual generation of rates and secondary yields for all available interactions can be done now with only a few lines of python code. This includes routines to copy the generated tables to the correct CRPropa directories, so they can be used in the same way as the default photon fields. An updated example explaining the consistent generation of custom photon target step-by-step is now included in the documentation\footnote{\url{https://crpropa.github.io/CRPropa3/pages/example\_notebooks/custom\_photonfield/custom-photon-field.html}}. Figure \ref{fig:customphotonfields} shows the spectral energy distributions of two custom photon field examples in comparison with the cosmic microwave background.

\section{New Physics Use Cases}
\label{sec:phys}

In this section, some examples of CRPropa simulations are presented to highlight different capabilities of the software.

\subsection{Diffuse Shock Acceleration with Stochastic Differential Equations}
\label{ssec:sdedsa}
Diffusive Shock Acceleration (DSA) can be simulated using the \texttt{DiffusionSDE} module based on Stochastic Differential Equations (SDEs). In the ensemble-averaged picture, DSA is described by adiabatic energy change. Using the SDE approach, change in momentum $p$ is given by
\begin{equation}
    \label{SDEII}
    \mathrm{d}p = - \frac{p}{3} \nabla \cdot \vec{u} \, \mathrm{d}t,
\end{equation}
with advection field $\vec{u}$. From eq.~(\ref{SDEII}) follows, that ideal shocks cannot be simulated. A planar one-dimensional shock with finite width can be approximated, e.g. by $u(x) \propto \tanh\left( x/L_{\mathrm{sh}}\right)$. In order to simulate an ideal shock, the shock width $L_{\mathrm{sh}}$ has to be small compared to the diffusion step length, but pseudo-particles also need to encounter the region of changing advection field to experience sufficient acceleration. For a detailed discussion on the choice of time step and shock width we refer to \cite{AerdkerEA23}. 
Injecting mono-energetic pseudo-particles of energy $E_0$ at the shock $x = 0$ and considering a stationary flux of particles to arrive at the shock, the time evolution of the spectrum can be approximated using the \texttt{TimeEvolutionObserver} (see \cite{CRPropa31, AerdkerEA23} for details).

\begin{figure}[htbp]
    \centering
    \includegraphics[width = 1\textwidth]{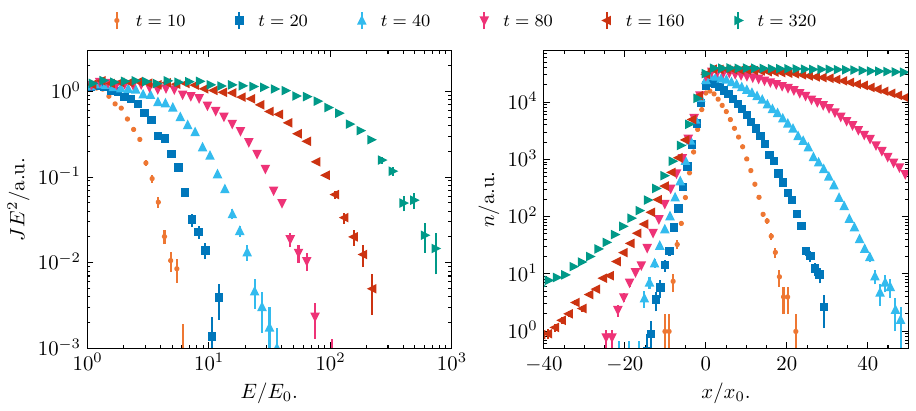}
    \caption{Time evolution of energy spectrum at a planar one-dimensional shock (left) and number density in acceleration region (right). $L_{\mathrm{sh}} = 0.02$, upstream speed $u_1 = 1$, compression $q = 4$ all units normalised $x_0 = v_0 t_0 = 1$. Diffusion is energy-dependent, $\kappa(E) = \kappa_0 (E/E_0)^{0.5}$ with $\kappa_0 = 1$. }
    \label{fig:spectra}
\end{figure}

Figure \ref{fig:spectra} shows the time evolution of the energy-spectrum at the downstream side of the shock and the number density in the acceleration region. Energy-dependent diffusion, $\kappa(E) = \kappa_0 (E/E_0)^{0.5}$ is considered. The compression ratio of the shock is $q = 4$, leading to a spectral slope $s = -2$ for an ideal shock. To increase statistics at high energies a new \texttt{CandidateSplitting} module was used, which will become available in a future release. Candidates are split in two copies once they cross energy-boundaries depending on the expected spectral slope. 

With CRPropa3.2 it is possible to simulate time-dependent DSA. Also acceleration at shocks with finite width, leading to steeper energy-spectra can be modelled. 

\subsection{Cosmic Ray Electron Transport in M51}
\label{ssec:M51}

The face-on galaxy M51 was recently used to demonstrate that three dimensional transport modelling helps to differentiate between advective and diffusive propagation \cite{Doerner23}. Radio observation hinted on an energy-independent diffusion and a radial dependence of the electron spectral index, which is hard to explain with purely advective or diffusive transport models. 

For faster computations the authors implemented an extension of CRPropa for the relevant energy losses (synchrotron radiation and inverse Compton scattering) as continuous loss approximation. These losses use the measured radial dependent magnetic field strength and a homogeneous energy density of the inter stellar radiation field as input parameters.
The observed star formation rate surface density has been used to draw the source positions and estimate the wind velocity. 

Based on these simulations it was concluded: 1) The diffusion seems to be energy independent, but 2) the scale height of the galaxy changes with radius from $z_0=3.2\,\mathrm{kpc}$ to $z_0=8.8\,\mathrm{kpc}$. Furthermore, it could be shown that earlier estimates of the wind velocity were approximately a factor five too large according to their model. 

\section{Summary and Outlook}
\label{sec:outlook}

We summarised new developments and extensions of the CRPropa simulation framework since its last major release version 3.2. Most of the updates are focused on better usability, e.g., by allowing for much simpler creation of custom photon fields, tracking the production of secondaries, or code software updates that improve the simulation in the background, such as the speed up for fixed propagation steps. 

These extensions will help to use CRPropa in a more flexible way, allowing to model, e.g., propagation and interactions in and close to the source of cosmic rays, test effects of non-standard redshift evolution of cosmic microwave background (CMB) models\footnote{This would be expected, e.g, for an SU(2) instead of a U(1) gauge theory (see, e.g., \cite{Hahn17}).}, on the diffuse gamma-ray flux, or include a consistent description of shock acceleration in ensemble-averaged transport models.

The plugins and examples demonstrate that a flexible simulation framework like CRPropa will be one of the key ingredients in upcoming multi-messenger models and help to improve our understanding of transport and interactions of cosmic rays and their secondaries better.

In the future, CRPropa should be extended to include, e.g., a model of momentum diffusion allowing to describe second-order Fermi acceleration within the SDE approach. Together with ongoing developments of nuclei-nuclei interaction this will open new opportunities to build accurate Galactic transport models.

Information on how to download the code, an online documentation, and examples of applications  (including some of those presented in this work) can be found in {\url{https://crpropa.desy.de/}}.

\section*{Acknowledgements}
RAB is funded by the ``la Caixa'' Foundation (ID 100010434) and the European Union's Horizon~2020 research and innovation program under the Marie Skłodowska-Curie grant agreement No~847648, fellowship code LCF/BQ/PI21/11830030. SA, JD, BE, KHK, LM, PR, LS, and JT acknowledge support from the Deutsche Forschungsgemeinschaft (DFG): this work was performed in the context of the DFG-funded Collaborative Research Center SFB1491 "Cosmic Interacting Matters - From Source to Signal", project no.\ 445052434 (SA, JD, BE, KHK, LM, PR, LS, JT). It was further supported by the project \textit{Multi-messenger probe of Cosmic Ray Origins (MICRO)}, project no.\ 445990517 (PR, LS, JT, KHK). The work of AS is supported by the Russian Science Foundation under grant no.~22-11-00063. KHK and GS acknowledge support by the BMBF Verbundforschung under grants 05A20PX1 and 05A20GU2. The work of PR was supported by a Gateway Fellowship. GS acknowledges support by the Deutsche Forschungsgemeinschaft (DFG, German Research Foundation) under Germany’s Excellence Strategy – EXC 2121 Quantum Universe – 390833306. AvV acknowledges support from the European Research Council (ERC) under the European Union’s Horizon 2020 research and innovation program (Grant No.~646623) and from Khalifa University's FSU-2022-025 grant. The work of PR was supported by a Gateway Fellowship.


\end{document}